\documentclass{article}

\usepackage{arxiv}

\usepackage{algorithmic}
\usepackage[utf8]{inputenc} % allow utf-8 input
\usepackage[T1]{fontenc}    % use 8-bit T1 fonts
\usepackage{hyperref}       % hyperlinks
\usepackage{url}            % simple URL typesetting
\usepackage{booktabs}       % professional-quality tables
\usepackage{amsfonts}       % blackboard math symbols
\usepackage{amsmath}
\usepackage{nicefrac}       % compact symbols for 1/2, etc.
\usepackage{microtype}      % microtypography
\usepackage{algorithm}
\usepackage{textcomp}
\usepackage{multirow}
\usepackage{numprint}
\usepackage[group-separator={,}]{siunitx}
\npthousandsep{,}\npthousandthpartsep{}\npdecimalsign{.}

\usepackage{graphicx}
\usepackage{subcaption}

\usepackage[normalem]{ulem}
\usepackage{xcolor}
\newcommand{\hl}[1]{{{\color{red}#1}}}

\newcommand{\CR}{$\mathsf{CR}$}
\newcommand{\BR}{$\mathsf{BR}$}
\newcommand{\GNN}{$\mathsf{GNN}$}
\newcommand{\CPU}{$\mathsf{CPU}$}
\newcommand{\GPU}{$\mathsf{GPU}$}
\newcommand{\DGL}{$\mathsf{DGL}$}

\title{Deep Graph Library Optimizations for Intel\textregistered~x86 Architecture}

\author{
Sasikanth Avancha \\
  Parallel Computing Lab, Intel Labs \\
  Intel Corporation \\
  Bangalore, India \\
  \texttt{sasikanth.avancha@intel.com} \\
   \And
 Vasimuddin Md.\\
  Parallel Computing Lab, Intel Labs\\
  Intel Corporation\\
  Bangalore, India \\
  \texttt{vasimuddin.md@intel.com} \\
  \And
 Sanchit Misra\\
  Parallel Computing Lab, Intel Labs\\
  Intel Corporation\\
  Bangalore, India \\
  \texttt{sanchit.misra@intel.com} \\
  \And
  Ramanarayan Mohanty\\
  Parallel Computing Lab, Intel Labs\\
  Intel Corporation\\
  Bangalore, India \\
  \texttt{ramanarayan.mohanty@intel.com}
}

\begin{document}
\maketitle

\begin{abstract}
The Deep Graph Library (\DGL{}) was designed as a tool to enable structure learning from graphs, by supporting a core abstraction for graphs, including the popular Graph Neural Networks (\GNN{}). \DGL{} contains implementations of all core graph operations for both the \CPU{} and \GPU{}. In this paper, we focus specifically on \CPU{} implementations and present performance analysis, optimizations and results across a set of \GNN{} applications using the latest version of \DGL{} (0.4.3). Across 7 applications, we achieve speed-ups ranging from $1.5\times$-$13\times$ over the baseline \CPU{} implementations. 
%We also compare \GNN{} application performance on Intel\textregistered~ Xeon\textregistered~ 8280 Platinum \CPU{} with 28 cores with published GPU data using a previous version of \DGL{} (0.3); we observe that V100 is ~3.6x faster for GCN running the full graph. 
\end{abstract}

\section{Introduction}
\label{sec:intro}

Graph Neural Networks (\GNN{})~\cite{hamilton17nips,Kipf:2016tc,velickovic2018graph,xu2018how,kipf18rgcn} are a very important class of Neural Network algorithms for learning the structure of large, population-scale graphs. Often, \GNN{}s are combined with traditional graph structure discovery algorithms via traversal (e.g., Breadth-First Search ($\mathsf{BFS}$), Depth-First Search ($\mathsf{DFS}$), RandomWalk) to achieve higher accuracy in learning their structure. Given a graph $G = (\mathcal{V}, \mathcal{E})$, the neural network formulation in \GNN{}s implies that, unlike graph traversal algorithms, they attempt to learn structure of $G$ via low-dimensional representations associated with $\mathcal{V}$ or $\mathcal{E}$ or both. \GNN{} algorithms, broadly, learn these representations in two parts: in feature vectors $F_{v}$ and/or $F_{e}$ associated with $\mathcal{V}$ and $\mathcal{E}$, respectively and a set of graph-wide, shared parameters $W$. Via a recursive process called Aggregation, \GNN{}s encode multi-hop neighborhood representations in $F_{v}$ and/or $F_{e}$. Depending upon the specific algorithm and the task (e.g., node classification, link prediction etc.), feature vectors aggregation precedes or succeeds a shallow neural network, typically consisting one or more linear transforms followed by a classification or regression model etc; some algorithms additionally employ a self-attention mechanism.

Given that aggregation is the core operation in all \GNN{} training and inference algorithms, our focus in this paper is on accelerating aggregation performance on Intel\textregistered~ Xeon\textregistered~ high-performance \CPU{}s. %, along with other primitives. 
Let $t = (u,v,e)$ be a tuple, where $e$ is the edge, and $u \in U$ and $v \in V$ are the source and destination vertices, respectively. Inherently, the aggregation operation involves {\em message passing} between any two entities in $t$.  \DGL{} implements the aggregation operation via two basic primitives: ${\tt send}(x)$ and ${\tt recv}(y, \oplus)$, where $x, y \in t$ and $\oplus$ is a reduction operation. 
We observed that \DGL{} fuses send and recv into a single primitive ${\tt fused\_sr}(x,y,\oplus)$ when aggregation consists of a simple arithmetic operation (as described in~\cite{wang2019dgl}); \DGL{} implements built-in fused aggregation primitive in such cases. 
%Across a range of applications in \DGL{} we observed that the inbuilt primitives are heavily used.

\DGL{} implements unary aggregation primitives (e.g. $\mathsf{copy\_u}$ and $\mathsf{copy\_e}$) which reduce a set of source features into the destination feature. In \DGL{} parlance, the unary aggregation primitive is called {\em Copy-Reduce}.
Similarly, \DGL{} implements binary aggregation primitives (e.g. $\mathsf{u\_mul\_e\_add\_v}$), which first perform an element-wise binary operation on two input feature vectors, and reducing the result into the destination feature via message passing. In \DGL{} parlance, the binary aggregation primitive is called {\em Binary-Reduce}. We discuss these primitives in greater detail in the next section.

As Wang et al. describe in~\cite{wang2019dgl}, the {\tt DGLGraph} interface hides the details of the graph data structure (e.g., $\mathsf{CSR}$ or $\mathsf{COO}$) from the programmer to enable better productivity. However, this implies that the application performance using \DGL{} for \GNN{} training and inference depends on how well the graph data structure and its associated operations have been optimized for the underlying architecture. Our analysis of various applications using \DGL{} revealed that the aggregation operation is implemented using sub-optimal primitives (such as serialization, explicit buffer copies prior to reduction), resulting in low performance on the Intel\textregistered~ Xeon\textregistered~ processor family. In this paper, we optimized the aggregate primitives in \DGL{} on \CPU{} and demonstrated the per epoch time speedups as high as $13\times$ on \DGL{} \GNN{} applications.

The rest of the paper is organized as follows. 
Section~\ref{sec:dglprimitives} describes the aggregate primitives used by \DGL{}. Section~\ref{sec:dglopt} describes the optimizations applied to binary-reduce and copy-reduce primitives. Section~\ref{sec:other_primitives} discusses primitives implemented in the PyTorch framework that impact GNN performance, and their optimizations.
In Section~\ref{sec:results}, we discuss the results of our optimizations and show performance improvements across various \GNN{} applications on Intel\textregistered~ Xeon\textregistered~ processors. Section~\ref{sec:conclusion} concludes the paper.
%\sm{I suggest we talk about the applications first and show their profile to make a case for optimization of certain primitives and then talk about how we optimized them}

\section{Aggregation Primitives}
\label{sec:dglprimitives}

%% Introduction to the section
%In this section, we describe the DGL aggregate primitives: binary-reduce and copy-reduce. We also %discuss various configurations of these primitives implemented in DGL.
Our understanding and analysis of the aggregation primitives in \DGL{} lead to the conclusion that these operations can be represented as a sequence of linear algebraic expressions involving node and edge features, along with the appropriate operators. The expression is sufficient to describe the aggregation over the complete graph or sub-graph upon which the operation executes.

\subsection{Binary-Reduce (\BR{})}
As discussed in section~\ref{sec:intro}, \BR{} consists of a sequence of two operations -- an element-wise binary operation between a pair of feature vectors, and an element-wise operation that {\it reduces} the intermediate feature vector into the output feature vector. When applied over the whole graph, the operands may be considered as multi-dimensional tensors representing nodes and edges.

Equation~\ref{eq:br} shows a mathematical representation of \BR{}, with operators $\otimes$ (element-wise binary operator) and $\oplus$ (element-wise reduction operator), and feature vector operands $x$, $y$ and $z$. $x$ and $y$ are inputs to $\otimes$, $\otimes(x,y)$ and $z$ are inputs to $\oplus$; the final result is in $z$.

\begin{gather}
BR(x, y, \otimes, \oplus, z): \oplus (\otimes(x, y), z), \label{eq:br} \\ \forall x, y, z \in G(\mathcal{V}, \mathcal{E}) \nonumber
\end{gather}

\begin{figure}[htbp]
\centering
\includegraphics{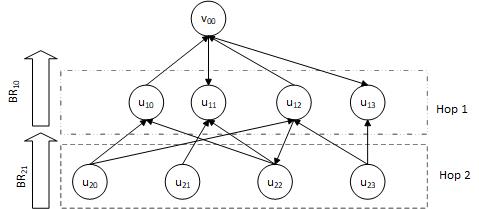}
\caption{Example directed subgraph showing aggregation direction. For node $v_{00}$, directed edges from all $u_{2x}$ to $u_{1x}$ to $v_{00}$ will be part of \BR{} operation.}
\label{fig:brcr}
\end{figure}

Figure~\ref{fig:brcr} shows an example subgraph induced on a (possibly) larger directed graph, and rooted at node $v_{00}$. In the example, all nodes labeled $u_{1x}$ are 1-hop neighbors of $v_{00}$; similarly nodes $u_{2x}$ are its 2-hop neighbors. Now, for example, $u\_mul\_e\_add\_v$ \BR{} between nodes $u_{20}$ and $u_{12}$, with feature vectors F$u_{20}$ and F$u_{12}$, respectively and edge-feature vector F$e_{2012}$ would have the following expression:

\begin{gather}
    BR_{21}(Fu_{20}, Fe_{2012}, \times, +, u_{12}): \nonumber \\  
    t \leftarrow Fu_{20} \times Fe_{2012} \nonumber \\
    u_{12} \leftarrow u_{12} + t \label{eq:bre}
\end{gather}

Further, we observe that nodes $u_{11}$ and $u_{13}$ will {\em not} be part of \BR{} on the subgraph rooted at $v_{00}$ because there are no edges {\em from} them {\em to} $v_{00}$.

If the size and shape of the input feature vectors are not equal, and if one of them has size $1$, then \BR{} {\em broadcasts} the smaller feature vector to the dimension of the larger one; in all other cases, \BR{} will fail to execute.

\subsection{Copy-Reduce (\CR{})}
\DGL{} implements the \CR{} operation separately from \BR{} as it is widely used in \GNN{} applications without any binary operation. Therefore, we view it as a special class of \BR{}. \CR{} takes only one input operand associated with the source (node or edge) and passes the feature vector as a message (i.e., {\it copies} it to the destination (node or edge), where it is reduced onto the latter. 

As shown in Equation~\ref{eq:cr}, \CR{} can be mathematically represented using \BR{} syntax with y replaced with $\mathsf{NULL}$ or $\phi$, resulting in $\otimes(x,y)$ becoming a {\it unary} operation {\tt copy(x)}:

%\DGL{} also implements the two special cases of \BR{} operations: (a) with binary operator and one of the two operand (i.e u or v) as null, (b) here,
%broadcast the input feature tensors (of u, v, e), when the feature tensors of input operands do not match (shape or dimension).
%In the former, \BR{} directly reduces the non-null input operand, we henceforth refer to this case as copy-reduce (\CR{}); while, in the latter case, usual \BR{} can be applied after the broadcasting, we henceforth refer to this case as binary-reduce-bcast.
%% \footnote{dgl use Binary-Reduce-Bcast term to for this case}.

\begin{gather}
BR(x, \phi, \otimes, \oplus, z): \oplus (\otimes(x, \phi), z) \nonumber \\
\otimes(x, \phi) = {\tt copy}(x) \nonumber \\
BR(x, \phi, \otimes, \oplus, z) \Rightarrow CR(x, {\tt copy}, \oplus, z) \nonumber \\
CR(x, {\tt copy}, \oplus, z): \oplus ({\tt copy}(x), z), \forall x, z \in G(\mathcal{V}, \mathcal{E}) \label{eq:cr}
\end{gather}

For example, in Figure~\ref{fig:brcr}, with sum (+) as reduction operation, \CR{} between $u_{10}$ and $v_{00}$ can be expressed as:

\begin{gather}
    CR_{21}(u_{10}, {\tt copy}, +, v_{00}): \nonumber \\
    t \leftarrow {\tt copy}(u_{10}) \nonumber \\
    v_{00} = v_{00} + {\tt copy}(u_{10}) \label{eq:cre}
\end{gather}
%% write down: what is our terminology for \CR{} and its two special cases.

\subsection{Configurations of Binary-Reduce and Copy-Reduce}
 Various configurations of \BR{} arise as a result of multiple candidates for each input operand and reduction destination. Here, we present the comprehensive list of configurations of \BR{} and \CR{} primitives implemented in \DGL{}. (Table~\ref{tab:configs}). 
 % The configurations are as follows, (here $\mathsf{bop}$ represents the binary operator ($\otimes$) and $\mathsf{rop}$ represents the reduction operation ($\oplus$)).
% We limit ourselves to arithmetic operators as binary operators: $\otimes \in \{ADD, SUB, MUL, DIV\}$, which we also found as common aggregators. Also, assuming the reduction destination to be either v or e, the list of primitives is as follows.

\noindent
\begin{table}[th]
\centering
\caption{Various configurations of \BR{} and \CR{} primitives inbuilt in \DGL{}. %{\em bop} and {\em rop} represent the binary and reduction operation respectively.
}
\label{tab:configs}
\begin{tabular}{c|c}
\hline
\textbf{\BR{}} & \textbf{\CR{}} \\ \hline 
$\mathsf{u\_\otimes\_v\_\oplus\_v}$,\hspace{10pt}  $\mathsf{v\_\otimes\_u\_\oplus\_v}$, & $\mathsf{u\_copy\_\oplus\_v}$,\hspace{10pt} $\mathsf{e\_copy\_\oplus\_v}$\\
$\mathsf{u\_\otimes\_v\_\oplus\_e}$, \hspace{10pt}$\mathsf{v\_\otimes\_u\_\oplus\_e}$, \\
$\mathsf{u\_\otimes\_e\_\oplus\_v}$,\hspace{10pt} $\mathsf{e\_\otimes\_u\_\oplus\_v}$, \\
$\mathsf{u\_\otimes\_e\_\oplus\_e}$,\hspace{10pt} $\mathsf{e\_\otimes\_u\_\oplus\_e}$, \\
$\mathsf{v\_\otimes\_e\_\oplus\_v}$,\hspace{10pt} $\mathsf{e\_\otimes\_v\_\oplus\_v}$, \\
$\mathsf{v\_\otimes\_e\_\oplus\_e}$,\hspace{10pt} $\mathsf{e\_\otimes\_v\_\oplus\_e}$ \\ 
\hline
\end{tabular}
\end{table}

\DGL{} has built-in support for a set of configurations in which $\otimes \in  \{\mathsf{add}, \mathsf{sub}, \mathsf{mul}, \mathsf{div}, \mathsf{dot}\}$  and $ \oplus \in \{\mathsf{add}, \mathsf{max}, \mathsf{min}, \mathsf{mul},\mathsf{div},\mathsf{copy} \}$. In practice, \DGL{} showed that these configuration are enough to support a large majority of applications. %% Give example where applications have used their own defined aggregation operators, lstm ??
Our evaluation showed that even with these simple operations, the \BR{} primitive executes for a majority of the run-time across various  applications (described in the Section~\ref{sec:results}). We profiled $7$ \GNN{} applications (total $8$ instances) and the \BR{} and \CR{} primitives used by them (Table~\ref{tab:cr_apps}).

%% \paragraph{Various Applications: \BR{} Configurations}:
%In our evaluation, we profiled total $11$  applications and optimized the various \BR{} and \CR{} configurations.% that runtime dominant in these applications. 
%Table~\ref{tab:cr_apps} shows the applications and the \BR{} and \CR{} configurations used by them. While, Table~\ref{tab:optimizations_charts} in the results section shows the results of optimizations on \BR{} primitive for these applications.

\begin{table}[th]
\centering
\caption{\GNN{} applications and \BR{} and \CR{} configurations used by them.}\label{tab:cr_apps}
\begin{tabular}{l|l|ll}
\hline
    & \textbf{Application}          & \textbf{\BR{} Configurations}     &  \\ \hline
 1. & GCN                  & ($\mathsf{u\_copy\_add\_v}$)  &   \\ \hline
 2. & GCN-Sampled          & ($\mathsf{u\_copy\_add\_v}$)  &   \\ \hline
 3. & GraphSAGE            & ($\mathsf{u\_copy\_add\_v}$)  &   \\ \hline
 4. & GraphSAGE-Sampled    & ($\mathsf{u\_copy\_add\_v}$)  &   \\ \hline
 5. & GCMC                 & ($\mathsf{u\_copy\_add\_v}$), ($\mathsf{u\_dot\_v\_add\_e}$) &   \\ \hline
 6. & Line Graph           & ($\mathsf{u\_copy\_add\_v}$) &   \\ \hline
 7. & Monet                & ($\mathsf{u\_mul\_e\_add\_v}$) &   \\ \hline
% 8. & MPNN                & ($\mathsf{u\_mul\_e\_add\_v}$) &   \\ \hline
 \multirow{4}{*}{8.}       & \multirow{4}{*}{GAT}                  & ($\mathsf{e\_copy\_add\_v}$), ($\mathsf{e\_copy\_max\_v})$, \\ 
                                                                  && ($\mathsf{u\_add\_v\_copy\_e}$), ($\mathsf{e\_sub\_v\_copy\_e}$), \\
                                                                  && ($\mathsf{e\_div\_v\_copy\_e}$), ($\mathsf{u\_mul\_e\_add\_v})$,\\
                                                                  && ($\mathsf{v\_mul\_e\_copy\_e}$) &   \\ \hline
 9. & RGCN-Hetero         & ($\mathsf{u\_copy\_add\_v}$) &   \\ \hline
 %11. & RGCN                &  & 
\end{tabular}
\end{table}

\subsection{Baseline Implementations of \BR{} and \CR{} in \DGL{}}
\label{sec:basecr}
The graph adjacency matrix in \DGL{} is in Compressed Sparse Row (CSR) format. The \CPU{} implementation first loads the features of $u$, $v$ and/or $e$, as required, for each row offset (representing the source node) and corresponding column indices (representing destination nodes). Using these feature vectors, it performs \BR{} or \CR{} for the tuple $(u, v, e)$. Specifically, to execute \CR{}, \DGL{} implements a {\em push} model. By {\em push}, we mean that in $\oplus({\tt copy}(x_k), z_{k-1})$ executes {\it from} hop $k$ {\t to} $k-1$, so $x \in$ hop $k$ and $z \in$ hop $k-1$.
To achieve good performance for \CR{} on the \CPU{}, the implementation parallelizes the loop over rows of the $\mathsf{CSR}$ matrix (i.e., the source nodes, $x_k$). Since \CR{} is an integral part of \BR{}, we first focus on the problems associated with parallel execution in \CR{}. As shown in Equation~\ref{eq:br}, the binary operation $\otimes$ is straightforward, executes before $\oplus$ and can be parallelized easily, with the result stored in some temporary feature $t$.

Algorithm~\ref{algo:push} describes the baseline push model in \DGL{}'s \CR{} implementation.

\begin{algorithm}[htbp]
\caption{Copy-Reduce: Push}
\label{algo:push}
\begin{algorithmic}[1]
\small
\FORALL{source nodes $u \in \mathcal{V}$ in parallel}
    \STATE copy\_u(u, out) \hspace{1.75in} \COMMENT {\bf \DGL{} function that copies source feature vector  to out}
    \FORALL{destination nodes $v \in N(u)$ in serial}
        \STATE $F_{v} \leftarrow F_{v} \oplus out$
    \ENDFOR
\ENDFOR
\end{algorithmic}                            
\end{algorithm}

When different nodes $u$ share neighbor $v$ and if the \CR{} destination is nodes $v$, then the \CR{} operation results in race condition among the threads. \DGL{} employs serialization using critical sections to resolve the race condition. The serialization significantly impacts the performance, leading to slower application run times. 

Also, the {\tt push} approach to \CR{} is {\it scatter-heavy} given that the graph adjacency matrix is more than 99.9\% sparse; this bounds \CR{} performance by memory access latency to randomly scattered destination node addresses in memory. We also observed via profiling and analysis that there is a potential reuse 
%(approximately $2\times$ - $2.5\times$) of the feature vectors because that is
proportional to the average node degree. However, the {\tt push} model fails to make use of this reuse because it simply scatters the feature vector to different addresses. This results in a significant amount of wasted memory bandwidth.

\section{Optimizing Aggregation Primitives}
\label{sec:dglopt}

\BR{} and \CR{} account for a majority of the run-time in \GNN{} applications. In this section, we describe techniques we have created to optimize their implementations within \DGL{}. As discussed in Section~\ref{sec:basecr}, achieving high performance for \CR{} is critical to \BR{} performance as well; it is also potentially harder to achieve. Therefore, we first focus on \CR{} optimizations.

\subsection{Copy-Reduce}
To avoid the problems associated with the default {\tt push} model, \DGL{} provides a way to {\tt pull} messages from nodes $u$ and reduce them into nodes $v$. Now, parallelizing the \CR{} operation by distributing $v$ across OpenMP threads will not result in collisions because only one thread \emph{owns} all feature vector vectors $F_{v}$ at each destination node $v$ and reduces each \emph{pulled} source feature vectors into $F_{v}$ ~(Algorithm~\ref{algo:pull}).

\begin{algorithm}[htbp]
\caption{Copy-Reduce: Pull}
\label{algo:pull}
\begin{algorithmic}[1]
\small
\FORALL{destination nodes $v \in \mathcal{V}$ in parallel}
    \STATE copy\_u(u, out)  \hspace{1.75in} \COMMENT {\bf \DGL{} function that copies source feature vector  to out}
    \FORALL{source nodes $u \in N(v)$ in parallel}
        \STATE $F_{v} \leftarrow F_{v} \oplus out$
    \ENDFOR
\ENDFOR
\end{algorithmic}                            
\end{algorithm}

While Algorithm~\ref{algo:pull} solves the collision problem of Algorithm~\ref{algo:push}, it still does not solve either the feature vector reuse problem (to reduce wasted memory bandwidth) or the memory latency problem due to random access pattern of source addresses. It turns from a {\it scatter-heavy} algorithm to a {\it gather-heavy} one. To solve the problems associated with both {\tt push} and {\tt pull}, we further optimize this algorithm and implement a variant of the Sparse-Dense Matrix Multiply operation that Wang et al.~\cite{wang2019dgl} allude to. 
%% Talk about MM formulation explicitly?

The neighborhood graph is represented as a sparse matrix, the adjacency matrix in CSR format. The dense matrix consists of the feature vectors $F_{u}$ or $F_e$ associated with source nodes $u$ or edges $e$, respectively. Algorithm~\ref{algo:cropt} shows the details of our optimized \CPU{} implementation of \CR{}, for $\mathsf{u\_copy\_add\_v}$ configuration.

\begin{algorithm}[htbp]
\caption{Copy-Reduce: Pull Optimized}
\label{algo:cropt}
\begin{algorithmic}[1]
\small
%\REQUIRE M = CSR.num\_rows
%\REQUIRE K = CSR.num\_cols, kb = block-size on K dimension
%\REQUIRE N = length($F_{u}$), nb = block-size on N dimension
\REQUIRE $A$ - Matrix of size $M \times K$ in CSR format
\REQUIRE $B$ - Dense matrix of size $K \times N$
\REQUIRE $C$ - Dense matrix of size $M \times N$
\REQUIRE Reduction-operator: $\oplus$
\REQUIRE $N$ = length($F_{u}$),  $kb$ = block-size on $K$ dimension, $nb$ = block-size on $N$ dimension 
\REQUIRE $C \leftarrow 0$
\FOR{$r \in 0,\dots,M-1$ in parallel}
    \FOR{$c \in 0,\dots,K-1$, step $kb$}
        \STATE $B[c,\dots,c+kb] \leftarrow {\tt RadixSort}(B[c,\dots,c+kb])$
        \FOR{$n \in 0,\dots,N-1$, step $nb$}
            \STATE $C[r][n] \leftarrow += B[c][n]$  \hspace{1.2in}\COMMENT{\textbf{C[r] and B[c] are N-wide vectors}}
        \ENDFOR
    \ENDFOR
\ENDFOR
\end{algorithmic}                            
\end{algorithm}

The critical part of this formulation is that the rows and columns of the sparse matrix A represent the destination (M) and source (K) nodes, respectively and the dense matrix B consists of the source node feature vectors $F_{u}$. Thus, the output matrix C consists of feature vectors of destination nodes $F_{v}, v \in {\tt Neighborhood}(u)$, reduced from multiple source nodes $F_{u}$. Given that A is an adjacency matrix in $\mathsf{CSR}$ format, each row (i.e., $v$) only consists of column indices of connected source nodes $u$. So, in effect, the {\it matrix multiply} operation is to select those rows (i.e., source nodes $u$) of feature vectors $F_{u}$ from B that reduce into rows (i.e., destination node $v$) feature vectors $F_{v}$ in C. 

To achieve high performance, Algorithm~\ref{algo:cropt} contains two primary optimizations:
\begin{enumerate}
    \item Parallelizes over rows of A (and C). This optimization is similar to that in Algorithm~\ref{algo:pull}, where threads own destination nodes, and thus, there is no collision problem that we observe in Algorithm~\ref{algo:push}.
    \item Takes advantage of the {\it reuse} present in the graph, and avoids random gathers by:
    \begin{enumerate}
        \item Blocking the K dimension of A and B, ensuring that all threads work on one block of ${\tt kb}$ source nodes at a time,
        \item Sort the block of rows in B according to row-id using Radix Sort, and
        \item Block the N dimension of B and C to process ${\tt nb}$ feature vector elements at a time
    \end{enumerate}
\end{enumerate}

Due to 2(a), any feature vector in B read by some thread $t$ could be in the L2 cache of the \CPU{} if/when some other thread $t'$ reads the same feature vector. Due to 2(b), accesses of source node feature vectors from DRAM are not completely random, but in ascending order of addresses - which should help reduce DRAM access latency. Due to 2(c), all threads work only on a block of C of size ${\tt M} \times {\tt nb}$ at a time, where ${\tt nb}$ is the block size. We use a value of ${\tt nb}$ such that the block of C stays in the Last Level Cache ($\mathsf{LLC}$) of the \CPU{} until it is completely processed.

\subsection{Binary-Reduce}
We focus now on optimizing the binary operation within \BR{}, applying the optimized Algorithm~\ref{algo:cropt} to handle the \CR{} part. %We describe \BR{} optimizations in
Algorithms~\ref{algo:bropt1},~\ref{algo:bropt2},~\ref{algo:bropt3} describe the optimized BR for different configurations of input and output operations.

Our optimizations consist of three major steps. 
\begin{enumerate}
    \item Of the two input operands, gather the features of the second operand corresponding to each instance of the first operand, as required by the binary operation. 
    \item Perform the element-wise binary operation ($\otimes$) on the two operands. \label{step2}
    \item Reduce the dense matrix generated using $\oplus$. If the reduction destination is a node, then apply \CR{} on the node feature matrix. If the reduction destination is an edge, copy the result of Step~\ref{step2} to the dense edge feature matrix.
\end{enumerate}

To clarify the usage of various \BR{} configurations, we have shown three algorithms: (Node, Node, Any), (Node, Edge, Any) and (Edge, Node, Any) in Algorithms~\ref{algo:bropt1}, \ref{algo:bropt2} and \ref{algo:bropt3}, respectively. 

In Algorithm~\ref{algo:bropt1}, for each source node $u$, we load feature $F_u$ and gather connected destination node features $F_v$ (line 4). Depending on whether the final destination of reduction or copy is $u$, $v$ or the edge between them $e$, lines 6, 8 and 11, scatter the result $F_u \otimes F_v$ to the node-feature matrix $V_f$ or $E_f$, respectively. 

In Algorithm~\ref{algo:bropt2}, the second operand is the edge incident on $u$; therefore, we must first obtain the edge index $e$ from the incidence matrix $E^T$, gather its feature $F_e$ from edge-feature matrix $E_f$ and then perform $\otimes$ followed by reduction or copy on lines 7, 9, or 11, respectively, corresponding to the final destination. 

In Algorithm~\ref{algo:bropt3}, the first operand is the set of all edges $E$; in line 2, we load each edge-feature $F_e$; the second operand is the set of nodes $V$ upon which $e$ is incident; therefore, in line 4, we gather node-features $F_u \forall u \in V$; again, depending on the final destination, we reduce or copy $F_e \otimes F_u$ to $V_f$ or $E_f$ in lines 6, 8 and 10, respectively.

%For example, consider $\mathsf{u\_mul\_e\_add\_v}$ configuration of \BR{}. All the source nodes $u \in U$ are divided among the threads. For each node $u$, get the edge index $e$ from the incidence matrix, gather edge-feature $F_e$ from edge-feature matrix $EF$, element-wise multiply it with node-feature $F_u$ and apply CR on the product to {\it scatter reduce} it into node-feature $F_v$.
%Each node $u$ scatters its features to all the edges $e$. At each edge $F_u$ and $F_e$ are element-wise multiplied and stored into intermediate dense matrix $D$ where each row $d$ represents $F_e^{\prime}$. Then, a \CR{} operation is performed on matrix $D$ as \CR{}(d, {\tt copy}, +, v).

\begin{algorithm}[t]
\caption{Binary-Reduce: (Node, Node, Any)}
\label{algo:bropt1}
\begin{algorithmic}[1]
\small
\REQUIRE Matrix $A$ of size $M \times K$ in CSR format
\REQUIRE Matrix $E$ of size  $M^2 \times K$ in CSR format (Incidence)
\REQUIRE Matrix $E^T$ is $K \times M^2$ in CSR format (Incidence)
\REQUIRE Feature matrix $V_f$ of size $M \times d$ 
\REQUIRE Feature matrix $E_f$ of size $M^2 \times d$
\REQUIRE Input operands: X (Nodes), Y (Nodes)
\REQUIRE Output operand: Z (Edges)
\REQUIRE Binary-operator: $\otimes$, Reduction-operator: $\oplus$
\FOR{ $u \in 0,\ldots, M-1$ in parallel }
    \STATE $F_u \leftarrow V_f[u]$
    \FOR{ $v$ in A[$u$] }
        \STATE $F_v \leftarrow V_f[v]$
        \IF {$Z = U$}   
        \STATE   $V_f[u] \leftarrow F_u \oplus (F_u \otimes F_v)$ \hspace{1.2in}\algorithmiccomment{\textbf{Reduction Destination: source nodes $u$}}
        \ELSIF{$Z = V$}   
         \STATE  $V_f[v] \leftarrow F_v \oplus (F_u \otimes F_v)$ \hspace{1.2in}\algorithmiccomment{\textbf{Reduction Destination: destination nodes $v$}}
        \ELSIF{$Z = E$}   
         \STATE  $e \leftarrow E^T[v]$ 
         \STATE  $E_f[e] \leftarrow F_u \otimes F_v$  \hspace{1.6in}\algorithmiccomment{\textbf{Copy Destination: Edges}}
        \ENDIF
    \ENDFOR
\ENDFOR
\end{algorithmic}                            
\end{algorithm}

\begin{algorithm}[t]
\caption{Binary-Reduce: (Node, Edge, Any)}
\label{algo:bropt2}
\begin{algorithmic}[1]
\small
\REQUIRE Matrix $A$ of size $M \times K$ in CSR format
\REQUIRE Matrix $E$ of size  $M^2 \times K$ in CSR format (Incidence)
\REQUIRE Matrix $E^T$ is $K \times M^2$ in CSR format (Incidence)
\REQUIRE Feature matrix $V_f$ of size $M \times d$ 
\REQUIRE Feature matrix $E_f$ of size $M^2 \times d$
\REQUIRE Input operands: X (Nodes), Y (Edges)
\REQUIRE Output operand: Z (Any)
\REQUIRE Binary-operator: $\otimes$, Reduction-operator: $\oplus$
\FOR{ $u \in 0,\ldots, M-1$ in parallel }
    \STATE $F_u \leftarrow V_f[u]$
    \FOR{ $v$ in A[$u$] }
        \STATE $e \leftarrow E^T[v] $
        \STATE $F_e \leftarrow E_f[e]$
        \IF {$Z = U$}
        \STATE    $V_f[u] \leftarrow F_u \oplus (F_{u} \otimes F_e)$ \hspace{1.2in}\algorithmiccomment{\textbf{Reduction Destination: source nodes $u$}}
        \ELSIF{$Z = V$}
         \STATE  $V_f[v] \leftarrow F_v \oplus (F_{u} \otimes F_e)$ \hspace{1.2in}\algorithmiccomment{\textbf{Reduction Destination: destination nodes $v$}}
        \ELSIF{$Z = E$}
         \STATE  $E_f[e] \leftarrow F_{u} \oplus F_e$  \hspace{1.6in}\algorithmiccomment{\textbf{Copy Destination: Edges}}
        \ENDIF
    \ENDFOR
\ENDFOR
\end{algorithmic}                            
\end{algorithm}

\begin{algorithm}[t]
\caption{Binary-Reduce: (Edge, Node, Any)}
\label{algo:bropt3}
\begin{algorithmic}[1]
\small
\REQUIRE Matrix $A$ of size $M \times K$ in CSR format
\REQUIRE Matrix $E$ of size  $M^2 \times K$ in CSR format (Incidence)
\REQUIRE Matrix $E^T$ is $K \times M^2$ in CSR format (Incidence)
\REQUIRE Feature matrix $V_f$ of size $M \times d$ 
\REQUIRE Feature matrix $E_f$ of size $M^2 \times d$
\REQUIRE Input operands: X (Edges), Y (Nodes)
\REQUIRE Output operand: Z (Any)
\REQUIRE Binary-operator: $\otimes$, Reduction-operator: $\oplus$
\FOR{ $e \in 0,\ldots, M^2-1$ in parallel }
    \STATE $F_e \leftarrow E_f[e]$
    \FOR{ $u$ in E[$e$] }
        \STATE $F_u \leftarrow V_f[u]$
        \IF {$Z = U$}
        \STATE    $V_f[u] \leftarrow F_u \oplus (F_e \otimes F_{u})$ \hspace{1.2in}\algorithmiccomment{\textbf{Reduction Destination: source nodes $u$}}
        \ELSIF{$Z = V$}
         \STATE  $V_f[v] \leftarrow F_u \oplus (F_e \otimes F_{u})$ \hspace{1.2in}\algorithmiccomment{\textbf{Reduction Destination: destination nodes $v$}}
        \ELSIF{$Z = E$}
         \STATE  $E_f[e] \leftarrow F_e \oplus F_{u}$ \hspace{1.6in}\algorithmiccomment{\textbf{Copy Destination: Edges}}
        \ENDIF
    \ENDFOR
\ENDFOR
\end{algorithmic}                            
\end{algorithm}

As can be seen,  Algorithm~\ref{algo:cropt} is critical for the performance of both \BR{} and \CR{} operations. The algorithm is designed and optimized for small input matrices, usually occurring in applications that sample and batch the input graph for processing. However, the algorithm, right now, is not fully optimized for large input matrices, usually occurring in applications processing full graph in non-batched mode. Thus, for applications with full graph processing we make use of {\tt mkl\_sparse\_?\_mm()} MKL matrix multiplication kernel.

\section{PyTorch Primitives}
\label{sec:other_primitives}
We used PyTorch as the backend to execute \DGL{} and the neural network functions, e.g., Linear layer. Our application profiles indicated that a number of PyTorch primitives execute sub-optimally on the \CPU{}. Of these, {\it BatchNorm1d} and {\it Embedding} accounted for a significant amount of run-time in the Line Graph Neural Network ($\mathsf{LGNN}$) application. 

BatchNorm1d did not have an implementation within $\mathsf{MKLDNN}$ for PyTorch; therefore, we created an optimized version in a PyTorch extension by parallelizing across the samples and vectorizing across features per sample. The Embedding primitive in PyTorch is similar to Copy-Reduce in terms of operations: gather a set of feature vectors using index vectors and copy them into destination vectors in the Forward pass; scatter-reduce the gradients of Embedding weights in the Backward pass. 

\section{Results}
\label{sec:results}

In this section, we demonstrate the performance benefits of optimized aggregation and other primitives in various \GNN{} applications implemented in \DGL{}.

\subsection{Applications}
\label{sec:applications}
We analyzed and optimized seven applications that are implemented using \DGL{} and available within the \DGL{} Github repository~\url{https://github.com/dmlc/dgl/}. We briefly discuss these applications.
\begin{itemize}
    \item GCN~\cite{Kipf:2016tc} is a semi-supervised learning approach on graph-structured data that applies the notion of convolutions on graphs. In each layer, it applies linear transforms to regularized node features and normalizes them before aggregation.
    \item GraphSAGE~\cite{hamilton17nips} is a general inductive framework that uses node features to generate node embeddings for data unseen by the network. For each node $u$, it aggregates neighbor $v$ features $F_{v}$ and concatenates the aggregated $F_{v}$ to $F_{u}$ before applying a linear transform.
    \item Relational GCN (R-GCN)~\cite{kipf18rgcn} is a \GNN{} that applies the GCN framework to relational graphs. For each node $u$, it first aggregates linearly transformed neighbor feature $F_{v}$ {\it under relation} $r$ with $F_{u}$ and then aggregates them across all relations $r \in R$.
    \item Line Graph Neural Network ($\mathsf{LGNN}$)~\cite{chen2018supervised} is an instance of a \GNN{} that employs both node feature aggregation as well as edge-feature aggregation. Thus, there are two sequential aggregation steps that make this application particularly suitable for our optimization.
    \item MoNet~\cite{Monti_2017_CVPR} is a general framework for applying GCN to replace previous methods of learning on non-Euclidean  spaces, such as Geodesic CNN and Anisotropic CNN. In the \DGL{} implementation, the core aggregation step is $\mathsf{u\_mul\_e\_add\_v}$ (transform node features multiplied by Gaussian weights on the edges) followed by a sum, mean of max operation on the resulting feature vectors.
    \item Graph Convolutional Matrix Completion (GC-MC)~\cite{berg2017graph} is a graph-based auto-encoder framework for matrix completion that uses GCN for recommender systems. In the \DGL{} implementation, the aggregation operation is $\mathsf{copy\_u(u, out)}$ followed by sum reduction.
    \item Graph Attention Networks (GAT) leverage masked self-attentional layers by stacking layers in which nodes attend over their neighborhoods’ features. In this paper, we analyze GAT performance as applied to life-sciences applications such as molecules property prediction.
   % \item Relational GCN Hetero %(RGCN-Hetero)~\cite{schlichtkrull2018modeling} is an extension %of GCN framework, that operates on local graph neighborhoods of %large relational data, specifically for link prediction and %entity classification tasks. In this work, individual graph %nodes are updated by gather operation and followed by %accumulation with shared parameters across the whole graph.

    %%\item Message Passing Neural Network (MPNN)~\cite{pmlr-v70-gilmer17a} is a framework that employs message passing of edge features for $T$ time steps and an update step over node features. One of its primary applications domains is quantum chemistry.
    
\end{itemize}
%%\subsection{Experimental Setup and Analysis}
\subsection{Experimental Evaluation}
\label{sec:exp}

%% [DONE: Todo: Update the figures]
\begin{figure}[t] %%[htbp]
    \centering
    \begin{subfigure}{.24\textwidth}
    \includegraphics[width=.98\linewidth]{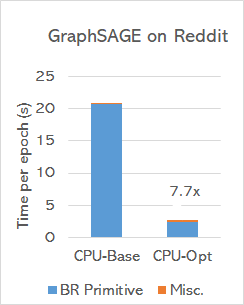}
    \end{subfigure}
    %\begin{subfigure}{.24\textwidth}
    %\includegraphics[width=.98\linewidth]{figures/graphsage_sampled_reddit.jpg}
    %\end{subfigure}
    \begin{subfigure}{.24\textwidth}
    \includegraphics[width=.98\linewidth]{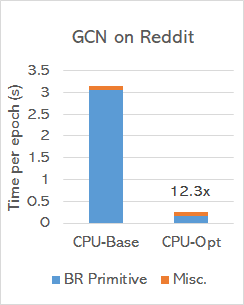}
    \end{subfigure}
    \begin{subfigure}{.24\textwidth}
    \includegraphics[width=.98\linewidth]{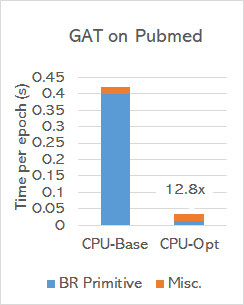}
    \end{subfigure}
    \begin{subfigure}{.24\textwidth}
    \includegraphics[width=.98\linewidth]{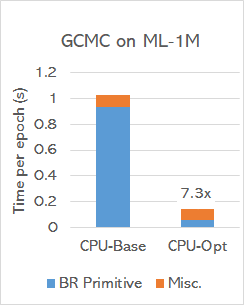}
    \end{subfigure}
    \begin{subfigure}{.24\textwidth}
    \includegraphics[width=.98\linewidth]{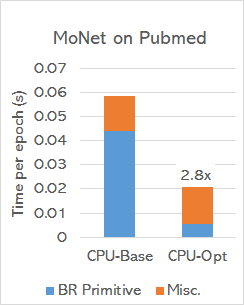}
    \end{subfigure}
    \begin{subfigure}{.24\textwidth}
    \includegraphics[width=.98\linewidth]{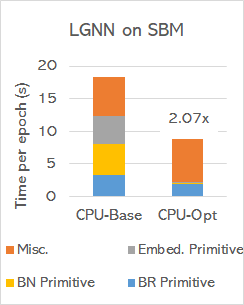}
    \end{subfigure}
    \begin{subfigure}{.24\textwidth}
    \includegraphics[width=.98\linewidth]{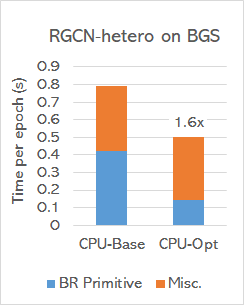}
    \end{subfigure}
    \begin{subfigure}{.24\textwidth}
    \includegraphics[width=.98\linewidth]{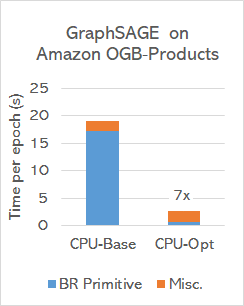}
    \end{subfigure}
    %\begin{subfigure}{.24\textwidth}
    %\includegraphics[width=.98\linewidth]{figures/graphsage_sampled_amazon.jpg}
    %\end{subfigure}
    \caption{Training performance comparison of \CPU{} optimizations against \CPU{} baseline on \DGL{} over seven different \GNN{} applications processing full graph in non-batched mode. The speedup by optimized code is mentioned on the top of the optimized bar. The Misc. is the run-time of all the remaining components. The performance numbers are averaged over 10 epochs, except for $\mathsf{LGNN}$, where we used 3 epochs. The datasets used for the experiments are mentioned in the charts. Here, \BR{} primitive represents the time for both \BR{} and \CR{}.} 
    \label{fig:dgl-0.4.3-cpu-performance}
\end{figure}

\begin{figure}[t] %%[htbp]
    \centering
    \begin{subfigure}{.24\textwidth}
    \includegraphics[width=.98\linewidth]{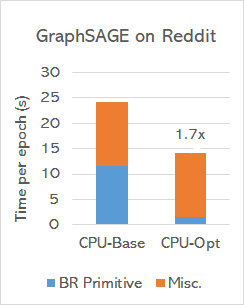}
    \end{subfigure}
    \begin{subfigure}{.24\textwidth}
    \includegraphics[width=.98\linewidth]{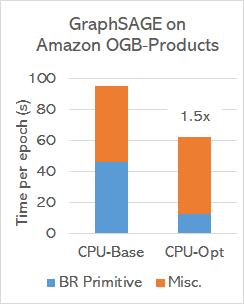}
    \end{subfigure}
    \caption{Training performance comparison of \CPU{} optimizations against \CPU{} baseline of GraphSAGE application with sampled graph processing on Amazon OGB-product and Reddit datasets. %over seven different \GNN{} applications. 
    The speedup by optimized code is mentioned on the top of the optimized bar. The Misc. is the runtime of all the remaining components. The performance numbers are averaged over 10 epochs. Here, \BR{} primitive represents the time for both \BR{} and \CR{}.
    } 
    \label{fig:dgl-0.4.3-cpu-performance-sampling}
\end{figure}

%% Replace \CR{} with \BR{}; here we represents \CR{} and \BR{} as one, \BR{} primitive.

\subsubsection{Experimental Setup}
\label{sec:exp-setup}
We performed all the experiments on  Intel\textregistered\  Xeon\textregistered~  8280 \CPU{} @2.70GHz with 28 cores (single socket), equipped with 98 {GB} of memory per socket. The peak bandwidth to DRAM on this machine is 128 GB/s. We used gcc v7.1.0 compiler for compiling \DGL{} and the backend PyTorch neural network framework from source code.

We used the latest release of \DGL{}v0.4.3 to demonstrate the performance enhancements due to our optimizations. 
%We also used \DGL{}v0.3+ to compare performance against published GPU numbers.
We used Pytorch v1.6.0-rc1 as the backend for all our experiments. %Specifically, we used Pytroch v1.4.0 with \DGL{}v0.3+, and Pytorch v1.5.0 with \DGL{}v0.4.3. 
All the applications execute with default parameter settings. We used Pytorch autograd profiler to profile the applications. 

\begin{table}[ht]
    \centering
    \caption{Benchmark graph dataset}
    \label{tab:dataset}
    \begin{tabular}{l|rrrr}
    \hline
Datasets     & \#nodes   & \#edges      & \#features & \#classes  \\
\hline
Pubmed       &  $\numprint{19717}$   &  $\numprint{44338}$    & $\numprint{500}$          &    $\numprint{3}$        \\
Reddit       & $\numprint{232965}$   & $\numprint{11606919}$   & $\numprint{602}$        & $\numprint{41}$         \\
% MovieLens-1M &           &              &    NA        &            \\
Amazon OGB-Products      & $\numprint{2449029}$ & $\numprint{123718280}$  & $\numprint{100}$        & $\numprint{47}$       \\
% SBM & - & - & - & - \\
BGS          &  $\numprint{44333}$   & $\numprint{227916}$      & $\numprint{103}$ (Relations) & $\numprint{2}$ \\
\hline
\end{tabular}
\end{table}
%% [Todo: We need to write SBM and ML-1M details in the text.]

Table~\ref{tab:dataset} shows the details of the datasets used in our experiments.
% In addition to the datasets mentioned in Table~\ref{tab:dataset},
Additionally, we used MovieLens-1M (ML-1M) dataset for GC-MC application and a synthetic dataset built using stochastic block model (SBM) for LGNN application. 
ML-1M is a benchmark dataset based on the user ratings for the movies; it consists of $ \numprint{6040}$ users, $\numprint{3706}$ movies, $ \numprint{1000209}$ ratings with rating levels $1,2,\ldots,5$. And, SBM is a synthetic dataset consists of random graph model with planted clusters. We used the default input parameters to generate the dataset.

%we used two other dataset types in our experiments, MovieLens-1M and Stochastic Block Model (SBM). The MovieLens-1M is a collaborative filtering benchmark dataset used in GC-MC application~\cite{berg2017graph}. This dataset is based on the user ratings for the movies. It consists of $ 6,040$ users,  \num{3706} movies, $\num[group-separator={,}]{1000209}$ ratings with rating levels $1,2,...,5$. However, SBM is a synthetic dataset consists of random graph model with planted clusters and used in LGNN application~\cite{chen2018supervised}. It is denoted by $SBM(n, p, q, C)$, where $n$~is the \#node, $p$ \& $q$~ are the probabilities of edge connecting any two vertices \& viceversa, and $C$~is the \#class.

\subsubsection{Performance Evaluation of \DGL{}}
%To demonstrate the performance of our optimizations, we chose seven popular applications from \DGL{} example applications (\url{dgl/examples/pytorch/}). 
We compared the  performance of optimized \DGL{} against the baseline (i.e non-optimized) \DGL{}. We ran all the seven applications with non-batched (full graph) processing; moreover, we also experimented with GraphSAGE with batched graph processing (sampled) (Figure~\ref{fig:dgl-0.4.3-cpu-performance} and Figure~\ref{fig:dgl-0.4.3-cpu-performance-sampling}). We used the biggest of the benchmark datasets provided in the \DGL{} for these applications. For GraphSAGE, we also show performance results for a  bigger dataset -- the Amazon ogb-products dataset from~\url{https://ogb.stanford.edu/docs/nodeprop/}.

Overall, for applications with non-batched processing, we observed a speedup of $1.6\times - 12.8\times$ on per epoch time over the baseline \DGL{} on the CPU; specifically, we observe \BR{} speedup between $1.72\times - 34\times$ per epoch time compared to the \DGL{} baseline across the seven application  (Figure~\ref{fig:dgl-0.4.3-cpu-performance}). 
Similarly, for GraphSAGE with batched processing, we see overall speedup of $1.5\times$-$1.7\times$  per epoch over \DGL{} baseline; specifically, we observe \BR{} speedup between $7.2\times - 10.6\times$ per epoch over \DGL{} baseline (Figure~\ref{fig:dgl-0.4.3-cpu-performance-sampling}). 
%In GraphSAGE with batch processing Algorithms~\ref{algo:cropt} saw gains as high as 20\% (not shown in figure). 
All our optimizations ensure the same accuracy as the baseline \DGL{}.

%For GraphSAGE sampled and LGNN, apart from \CR{}, other primitives consumes a sizable portion of the runtime. In our investigation we found that all these other primitives have definition in the backend framework \emph{i.e} Pytorch. 
%% should we talk about \DGL{} overheads?
%% Accuracy, geomean, other primitives, gcn-sampled

Our optimizations of BatchNorm1d and Embedding PyTorch primitives (in LGNN application) resulted in $13\times$ and $76\times$ respectively. Together with these three optimized primitives optimized LGNN achieves $2\times$ speedup over baseline.

The Misc. portion of the runtimes in Figure~\ref{fig:dgl-0.4.3-cpu-performance} is majorly contributed by other primitives -- due to Pytorch framework -- plus some \DGL{} framework overheads.
These PyTorch primitives can be optimized on similar lines as 1D Native Batch Norm and Embedding primitives. %% Need to write no small primitives in orange bar.

\section{Conclusions}
\label{sec:conclusion}
Aggregation operations are critical to Graph Neural Network applications functionality. Via extensive application profiling and analysis of their implementations in the popular \DGL{}, we observed that aggregation primitives account for a majority of the run-time across applications. The Binary-Reduce abstraction in \DGL{} is the main aggregate operation. It is a memory-intensive operation with element-wise operations being the only compute; therefore, on CPU, the performance of this primitive is bound by the available memory-bandwidth. We optimized the sparse-dense matrix multiplication formulation of binary-reduce (and its special case, copy-reduce). We have demonstrated the benefits of the optimizations across a range of \GNN{} applications in \DGL{}. 

%% Todo: Concluding statement about GNN?

\newpage
\bibliographystyle{unsrt}  
\bibliography{references}

\end{document}